\documentclass[twocolumn]{aastex62}
\usepackage{amsmath}
\usepackage[varg]{txfonts}
\usepackage{natbib}
\bibpunct{(}{)}{;}{a}{}{,}
\usepackage{xspace}
\usepackage{graphicx}
\usepackage{enumitem}

\hypersetup{linkcolor=red,citecolor=cyan,filecolor=green,urlcolor=magenta}

\newcommand{\Kij}{\mbox{$K^{i}_{j}$}\xspace}
\newcommand{\Knilone}{\mbox{$K^{\rm 1}_{\rm 0}$}\xspace}
\newcommand{\Ktwoone}{\mbox{$K^{\rm 2}_{\rm 1}$}\xspace}
\newcommand{\Kthreetwo}{\mbox{$K^{\rm 3}_{\rm 2}$}\xspace}
\newcommand{\Kfourthree}{\mbox{$K^{\rm 4}_{\rm 3}$}\xspace}

\newcommand{\nhhh}{\mbox{NH$_3$}\xspace}
\newcommand{\nnhp}{\mbox{N$_2$H$^+$}\xspace}

\newcommand{\ammo}{\mbox{$\mathrm{NH_{3}}$}\xspace}
\newcommand{\kms}{\mbox{$\mathrm{km~s^{-1}}$}\xspace}
\newcommand{\vlsr}{\mbox{$v_{\mathrm{lsr}}$}\xspace}
\newcommand{\Vlsr}{\mbox{$V_{\mathrm{lsr}}$}\xspace}

\received{January 24, 2019}
\accepted{March 16, 2020}
\submitjournal{ApJL}

\shorttitle{Spectral multiplicity in NGC1333}
\shortauthors{Sokolov et al.}

\begin{document}

\title{Probabilistic detection of spectral line components}

\correspondingauthor{Vlas Sokolov}
\email{vlas-sokolov@mpg-alumni.de}

\author[0000-0002-5327-4289]{Vlas Sokolov}
\affil{Max Planck Institute for Extraterrestrial Physics, Gie{\ss}enbachstra{\ss}se 1, D-85748 Garching bei M{\"u}nchen, Germany}

\author[0000-0002-3972-1978]{Jaime E. Pineda}
\affiliation{Max Planck Institute for Extraterrestrial Physics, Gie{\ss}enbachstra{\ss}se 1, D-85748 Garching bei M{\"u}nchen, Germany}

\author[0000-0003-0426-6634]{Johannes Buchner}
\affiliation{Max Planck Institute for Extraterrestrial Physics, Gie{\ss}enbachstra{\ss}se 1, D-85748 Garching bei M{\"u}nchen, Germany}

\affiliation{Instituto de Astrofísica, Facultad de F\'isica, Pontificia Universidad Cat\'olica de Chile, Casilla 306, Santiago 22, Chile}

\affiliation{Millennium Institute of Astrophysics, Vicu\~na. MacKenna 4860, 7820436 Macul, Santiago, Chile}

\author[0000-0003-1481-7911]{Paola Caselli}
\affil{Max Planck Institute for Extraterrestrial Physics, Gie{\ss}enbachstra{\ss}se 1, D-85748 Garching bei M{\"u}nchen, Germany}

\begin{abstract}
Resolved kinematical information, such as from molecular gas in star forming regions, is obtained from spectral line observations.
However, these observations often contain multiple line-of-sight components, making estimates harder to obtain and interpret.
We present a fully automatic method that determines the number of components along the line of sight, or the spectral multiplicity, through Bayesian model selection.
The underlying open-source framework, based on nested sampling and conventional spectral line modeling, is tested using the large area ammonia maps of NGC~1333 in Perseus molecular cloud obtained by the Green Bank Ammonia Survey (GAS).
Compared to classic approaches, the presented method constrains velocities and velocity dispersions in a larger area.
In addition, we find \replaced{multiple components in 10.6\% of the emission-bearing pixels in the map, yet these second components do not substantially change the velocity dispersion distribution obtained with a single fit component.}{that the velocity dispersion distribution among multiple components did not change  substantially from that of a single fit component analysis of the GAS data.}
These results showcase the power and relative ease of the fitting and model selection method, which makes it a unique tool to extract maximum information from complex spectral data.
\end{abstract}


\section{Introduction} \label{sec:intro}

Spectral line observations provide a unique insight into the kinematics of astrophysical sources.
However, the underlying physical complexity of the latter often results in observed spectra bearing a signature of multiple independent components with distinct radial velocities.
In particular, a line-of-sight superposition of material with different radial velocities emitting in optically thin regime will naturally result in a spectral profile with multiple peaks, as seen in both Galactic \citep[e.g.,][]{busquet+2013, tanaka+2013} and extragalactic \citep[e.g.,][]{koch+2018} sources, as well as in molecular cloud simulations \citep{clarke+2018}. The observational signature of the multiple components, as well as the underlying number of the line-of-sight velocity components, shall be referred to as \textit{spectral multiplicity} henceforth, and is the topic addressed in this work.

Despite the widespread occurrence of spectral multiplicity, the issue has yet to be addressed in a statistically sound manner.
Traditionally, spectral multiplicity is often assessed by eye \citep[e.g.,][]{rosolowsky+2008-nh3,Pineda+2010,Pineda+2011, hacar+tafalla2011, beuther+2015,Pineda+2015, hacar+2016-musca, pon+2016, Monsch+2018, Barnes+2018}.
Recently, the drastically increasing number of spectra that can be delivered by the contemporary instruments has led to the development of a semi-automatic approach, where a by-eye fit to an averaged spectrum is propagated into individual spectra within the averaged area \citep{hacar+2013, henshaw+2013, hacar+2017, henshaw+2016, hacar+2018}, and the individual components are judged against heuristic criteria.
More recent works have used information criteria to prevent over-fitting, supplementing the conventional methods above \citep[such as AIC in][]{henshaw+2016,Chen+2020}.

However, currently employed methods to determine spectral multiplicity have their shortcomings. 
The commonly used approach of evaluating the number of spectral components by eye or by imposing detection heuristics does not have a statistically sound foundation.
Furthermore, forward-fitting multiple components in this fashion is invalidated by the sheer number of spectra in contemporary datasets, where tens or hundreds of thousand spectra can comprise a single observational result. Moreover, as most current approaches employ local, and not global, regression methods, the multitude of spectra analyzed in contemporary datasets creates yet another complication where the output optimal parameters are highly dependent on the initial conditions of the fit.
\added{More sophisticated frameworks such as GaussPy+ \citep{riener+2019}, while providing a fully-automated way to separate Gaussian spectral components, come with limitations of their own, such as not supporting complex hyperfine structures and absorption line profiles in case of GaussPy+.}
It is for these reasons that an automated, global minimization framework, complete with statistically robust selection of spectral multiplicity, is sorely needed.

In this study, we present the results of a Bayesian model selection for measuring the spectral multiplicity on the Green Bank Ammonia Survey \citep[GAS,][]{friesen+pineda+2017} data for NGC1333 in the Perseus molecular cloud, through the use of the nested sampling algorithm \citep{skilling2004}. The framework we present requires minimal human interaction, delivers full sampling over the free parameter space and allows well-defined detection criteria. By treating each spectrum in the spectral cube independently, we avoid biasing neighbours.

\section{Method summary}

\begin{figure*}
    \centering
    \includegraphics[width=0.49\textwidth]{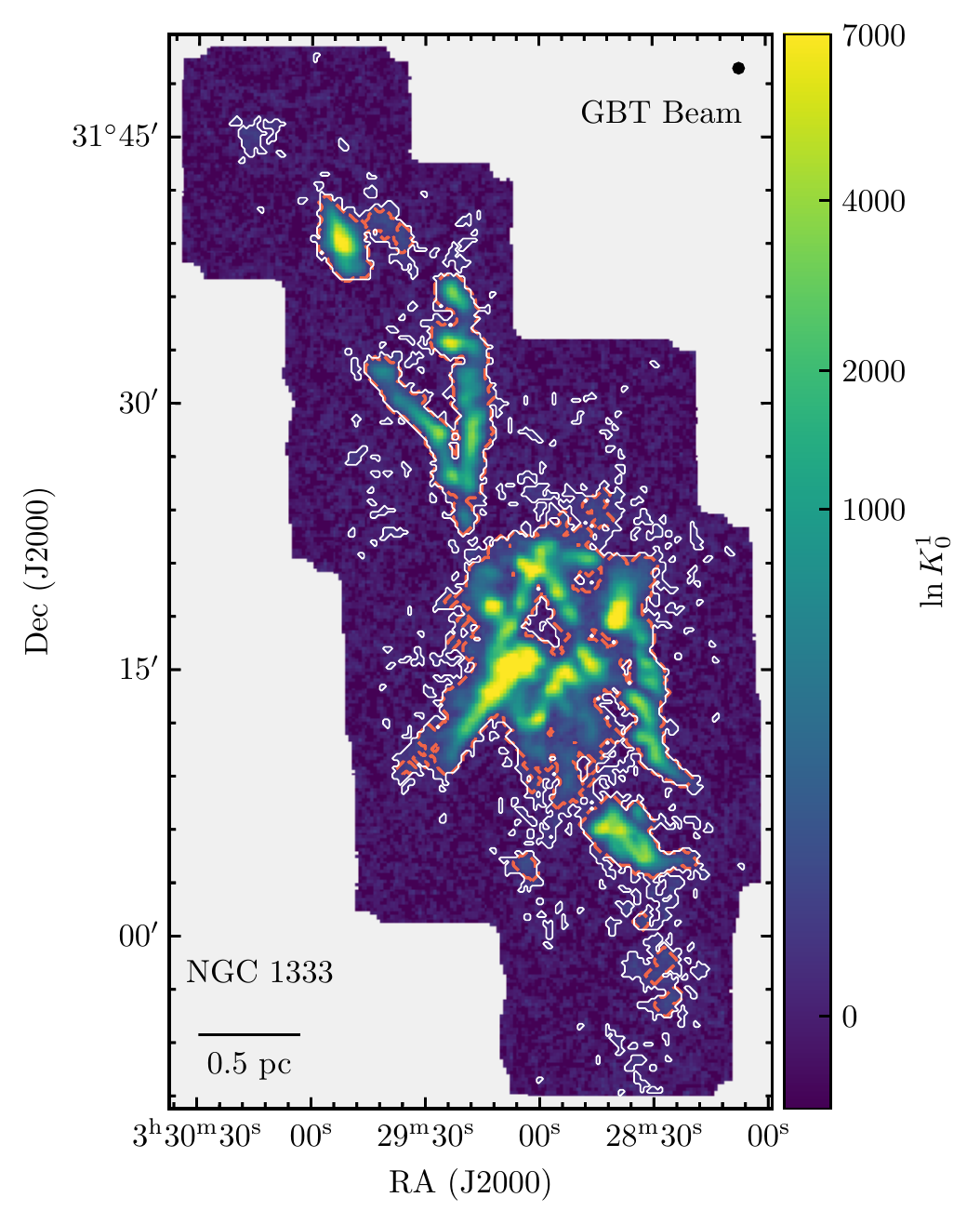}
    \includegraphics[width=0.49\textwidth]{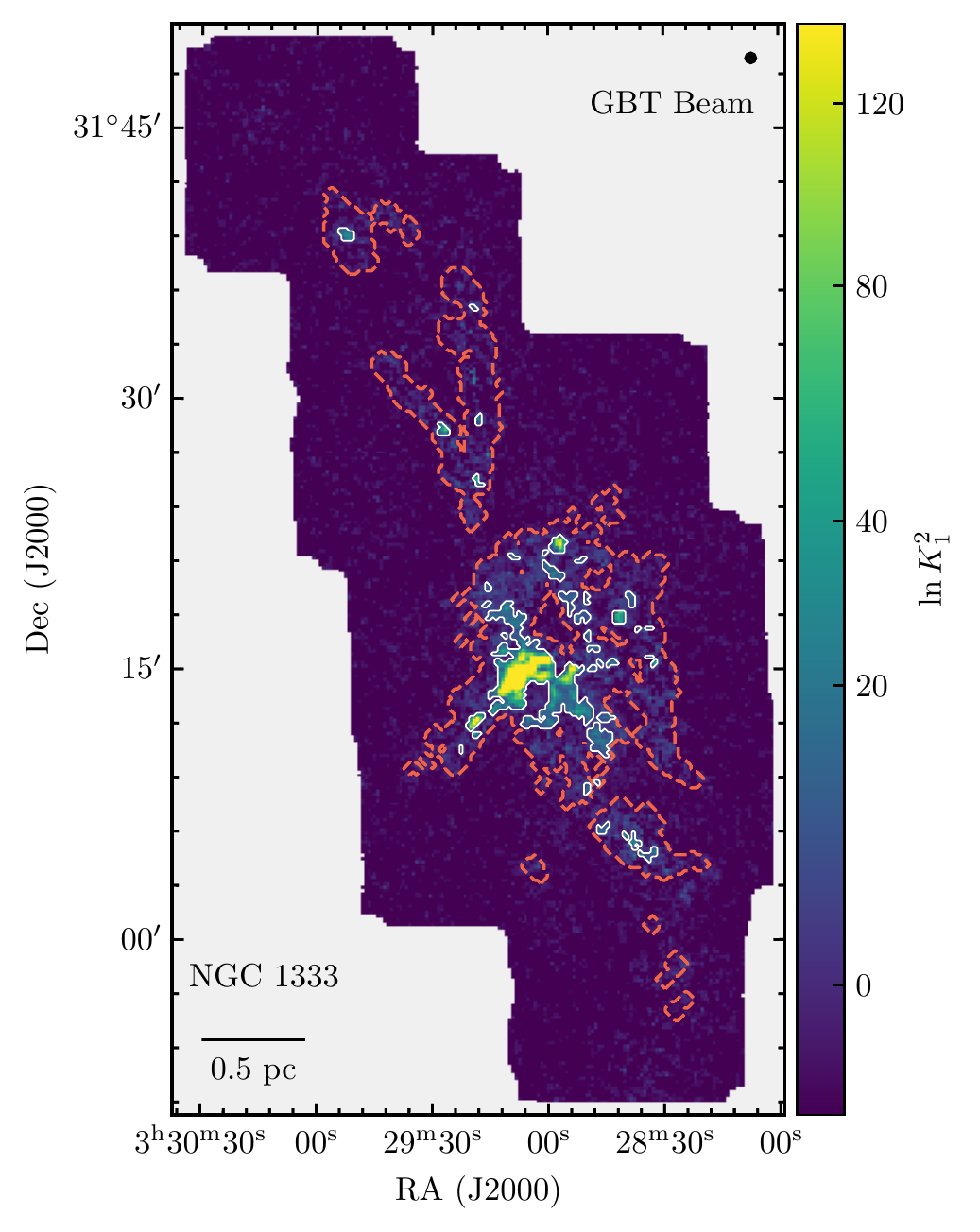}
    \caption{The odds ratios (Bayes factors) for detecting one (left, \Knilone) and two (right, \Ktwoone) spectral line components. Both maps are overplotted with the heuristical decision boundary of $\ln \Kij = 5$ (solid while lines) and the previous quality assessment for significant single spectra detection \citep[dotted red lines,][]{friesen+pineda+2017}. To highlight only spatially continuous spectral component detections, we have applied small feature removal on to the contours, only considering continuous regions of nine or more pixels.
    Scale bar and beam are shown in the bottom left and top right corners, respectively.}
    \label{fig:lnkmaps}
\end{figure*}

{Here we explain the classical method for line fitting and our new method.}

\subsection{Classical line fitting}

To model spectral lines, an observed spectrum is assumed to contain signal from an astrophysical source and a known Gaussian noise contribution with known amplitude $\sigma$. Fitting minimizes the $\chi^2$ statistic, defined as: 
\begin{equation*}
    \chi^2(\theta) = \sum_{i=1}^N \frac{(y_i - \mathcal{M}(\theta, i))^2}{\sigma^2}~,
\end{equation*}
where  $y_i$ is the observed signal amplitude at spectral channel $i$ (of $N$) and $\mathcal{M}(\theta, i)$ is the model of the astrophysical source dependent on some source parameters $\theta$ \citep[following][]{friesen+pineda+2017} that are varied during the fitting process. The square root of the Gaussian noise variance, $\sigma$, is taken from the GAS DR1 for our ammonia application.

The $\chi^2$ statistic originates from assuming a Gaussian likelihood valid across the spectral range, which can be written as 
\begin{equation}\label{eq: like}
\mathcal{L}(D|\theta) = \prod_{i=1}^{N} \frac{1}{\sqrt{2 \pi \sigma^2}} \times \exp{\left[-\frac{1}{2 \sigma^2} (y_i - \mathcal{M}(\theta, i))^2\right]}
\end{equation}
or, dropping constants, as 
\begin{equation}\label{eq: loglike}
\log \mathcal{L}(D|\theta) = \frac{1}{2} \chi^2~.
\end{equation}
Simple minimization algorithms have difficulties identifying globally best parameters in complex models $\mathcal{M}(\theta, i)$ when the signal is faint. There may be multiple local optima, or no well-defined optimum. Even if the fit succeeds, quantifying whether a more complex model is better than a simpler model can be difficult \citep{protassov+2002}. In the situation of deciding how many of $k$ components are justified by the data, the model of $k$ components contains the $(k-1)$ component model at the border of the parameter space (component amplitude is zero). However, just such border situations are not permitted by model comparison methods based on F-tests, likelihood ratio tests or more generally those relying on Wilks' theorem.

\subsection{Bayesian framework}

Bayesian model selection with nested sampling solves these problems elegantly. Nested sampling \citep{skilling2004} is a global parameter space exploration algorithm \added{(as opposed to local optimization in classical frameworks)}, which allows both parameter estimation and model comparison, even in low-signal data. \textit{Parameter estimation} gives the ranges of parameters $\theta$ that are probable given the data. To obtain \replaced{at the}{all the posterior} probability densities (and not just a \added{global maximum of the} likelihood function), we need to define prior densities over the parameter space. With these, we can use the likelihood to reweigh the prior probability density $\pi(\theta)$ to a posterior probability density $P(\theta|D)$ with Bayes' theorem:
\begin{equation*}
P(\theta|D) \propto \pi(\theta) \times \cal{L}(D|\theta),
\end{equation*}
\noindent \added{which can be used both for point estimation (i.e. finding the best-fit parameter values) as well as for uncertainty analysis (e.g. based on the posterior spread around the point estimate).}

In our application we use uninformative (flat) priors. The prior $\pi(\theta)$  defines the free parameters of the line modelling, which are listed below. Unless specified otherwise, the value ranges indicate uniform priors. 

\begin{enumerate}
    \item Gas kinetic temperature, ranging from CMB temperature to 25 K.
    \item Excitation temperature, ranging from CMB temperature to 25 K.
    \item Velocity dispersion of the line, ranging from 0.05 to 2 \kms.
    \item Logarithm of the ammonia column density, ranging from 12 to 15, probing H$_2$ column densities of $10^{20}$ to $10^{23}$ cm$^{-2}$ for typical ammonia abundances of $10^{-8}$.
    \item The mean \vlsr of all velocity components, ranging from 3 to 10 \kms.
    \item Velocity separation between the closest components, ranging from 0.2 to 3 \kms. 
\end{enumerate}

All the parameter priors, except for the mean \vlsr, are treated independently for each velocity component. The parameters 5 and 6 can be shown to linearly transform into a set of centroid velocities for each velocity component.
The priors above were chosen to be broad enough to encompass all typical physical conditions of the low-mass molecular clouds \citep[cf.][for overall distributions of the free parameters]{friesen+pineda+2017}.

\textit{Model comparison} evaluates whether to prefer one model over another. We first estimate the likelihood integral $Z$, called the Bayesian evidence, marginalised (integrated) over the entire parameter space,
\begin{equation*}
Z = \int p(\theta) \mathcal{L}(\theta) d\theta~.    
\end{equation*}

If we have a set of equally probable models under consideration, the Bayes factor \Kij  gives the odds ratio of $\mathcal{M}_i$ model over model $\mathcal{M}_j$:
The Bayes factor, \Kij, gives the odds ratio of $\mathcal{M}_i$ model over model $\mathcal{M}_j$:
\begin{equation}
    \Kij = \frac{\mathrm{P}(\mathcal{M}_j)Z_i}{\mathrm{P}(\mathcal{M}_i)Z_j} = Z_i / Z_j,
\end{equation}

\noindent where we implicitly assume that the competing models are equally likely \textit{a priori}, i.e., $\mathrm{P}(\mathcal{M}_i) = \mathrm{P}(\mathcal{M}_j)$.

Model comparison can also be used to quantify how strongly the data support the presence of a spectral line in the observed spectrum. A special case occurs when only noise is considered (i.e., $\mathcal{M} = 0$ in Eq. \ref{eq: like}). If the noise amplitude is known, the corresponding evidence $Z_0$ can be derived analytically \citep[see e.g.,][for a similar approach]{buchner2017}.

For our model selection purposes, we consider three models: one yielding a noise-only spectrum (with analytical evidence $Z_0$ above), the model where the data is a sum of noise and one spectral components (its evidence $Z_1$, sampled by MultiNest), and the two-component-model (with Bayesian evidence denoted as $Z_2$). Consequently, we will refer to two Bayes factors of interest in this study: \Knilone, denoting the odds of a spectrum being present in the data, and \Ktwoone, for the odds of two component model prevailing over single-component one.

We use the popular MultiNest \citep{feroz+hobson2008, feroz+2009} implementation of nested sampling through the PyMultiNest \citep{buchner+2014} Python interface. This has been widely used in the past to perform model selection for astrophysical spectra \cite[e.g.,][]{bernardi+2016, feldmeier-krause+2017, lavie+2017, baronchelli+2018}. 
The likelihood function was sampled through a wrapper to \verb+pyspeckit+, specifically designed to perform nested sampling of spectral cubes\footnote{\url{https://github.com/vlas-sokolov/pyspecnest}}.
The results of this study, as well as the code to reproduce them fully, are publicly available\footnote{\url{https://github.com/vlas-sokolov/bayesian-ngc1333}}.

\section{Data}

We demonstrate our technique on GAS DR1 \citep{friesen+pineda+2017} data in the NGC1333 region. 
These data cover $\approx 1222$\,arcmin$^{2}$ in the sky of the young embedded cluster in the nearby Perseus molecular cloud.
These data were observed using the On-The-Fly mapping technique 
with the K-Band Focal Plane Array (KFPA) at the Green Bank Observatory.
The spectral resolution of these data is 5.7\,kHz, or $\approx$0.07\,\kms at the frequency of these lines, 
and a beam FWHM of 32\arcsec.
The typical rms of the cube is $\approx$0.11\,K in the Main Beam scale.
We focus on \ammo~(1,1) and (2,2), at rest frequencies of 23.6944955 and 23.7226336\,GHz, respectively. 

The numerous hyperfine sub-components per line, high spectral resolution and sensitivity of the observations allow an accurate determination of the gas kinematics.
The initial GAS DR1 \citep{friesen+pineda+2017} fitted a single component fit using \verb+pyspecfit+, however acknowledged the presence of multiple components in a small fraction of the data.
In the same region, other works have shown a small fraction of multiple components along the line-of-sight and fit then multiple components by hand \citep[e.g.,][]{hacar+2017}.

\section{Results and Discussion} \label{sec:results}

We fitted each pixels spectrum separately with zero, one and two component models. We present the odds of detecting one and two components in Fig. \ref{fig:lnkmaps}. We emphasize that the sampling routine does not use knowledge about nearby pixels, except for the inherent correlation 
between nearby spectra due to pixel sizes smaller than beam size, and that we have used the same sampling setup for every spectrum sampled.

The odds ratios for detecting a single component are shown on left panel, overlaid with white contours of a conventional ($\ln \Knilone = 5$, see \S \ref{ssec:kcut}) cut. 
The red contours, marking the stringent detection criteria imposed by the GAS DR1 quality assessment \citep{friesen+pineda+2017}, are fully enclosed within the signal detected in our method. Additional regions emitting ammonia extend beyond the DR1 detection range, indicating that our probabilistic framework is sensitive to fainter emission than the heuristic quality cuts imposed in \cite{friesen+pineda+2017}.

\begin{figure}
    \centering
    \includegraphics[width=\columnwidth]{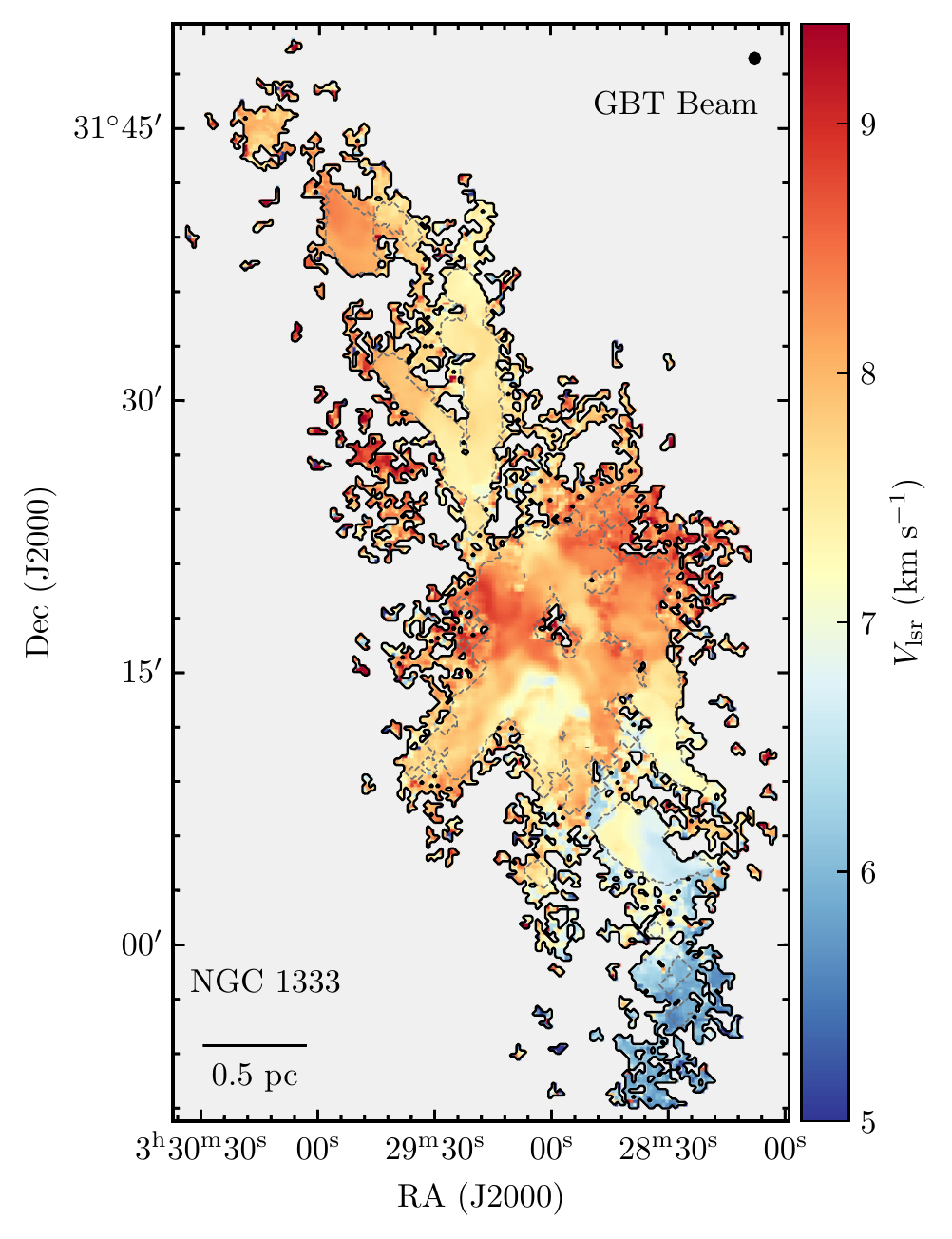}
    \caption{Map of maximum likelihood estimates for the centroid velocities \Vlsr, overlaid with $\Knilone=1$ contours. The extended spatial region shown (relative to $\Knilone \approx 100$ contours on Fig. \ref{fig:lnkmaps}) is showing the relative ease of constraining the centroid velocity parameter. 
    }
    \label{fig:mlexoff}
\end{figure}

The right panel of Fig.~\ref{fig:lnkmaps} shows the same Bayes factor cut ($\ln \Ktwoone = 5$) for model selection of detecting two components in white. The strongest evidence for the secondary component is found towards the eastern side of the main group of filaments, but isolated traces of secondary components can be found throughout the NGC~1333 region, beyond the regions where multiple components have been fit by hand in previous studies.

\subsection{Comparisons with previous work}

In Fig. \ref{fig:mlexoff} we plot the Maximum Likelihood Estimate (MLE) values of the centroid velocities at each pixel passing the decision boundary criterion above. The apparent lack of scatter in nearby pixels' values, even in the regions with faint emission (beyond the DR1 detection boundary) illustrates the relative ease in constraining the cloud kinematics.

\begin{figure}[t!]
    \centering
    \includegraphics[width=\columnwidth]{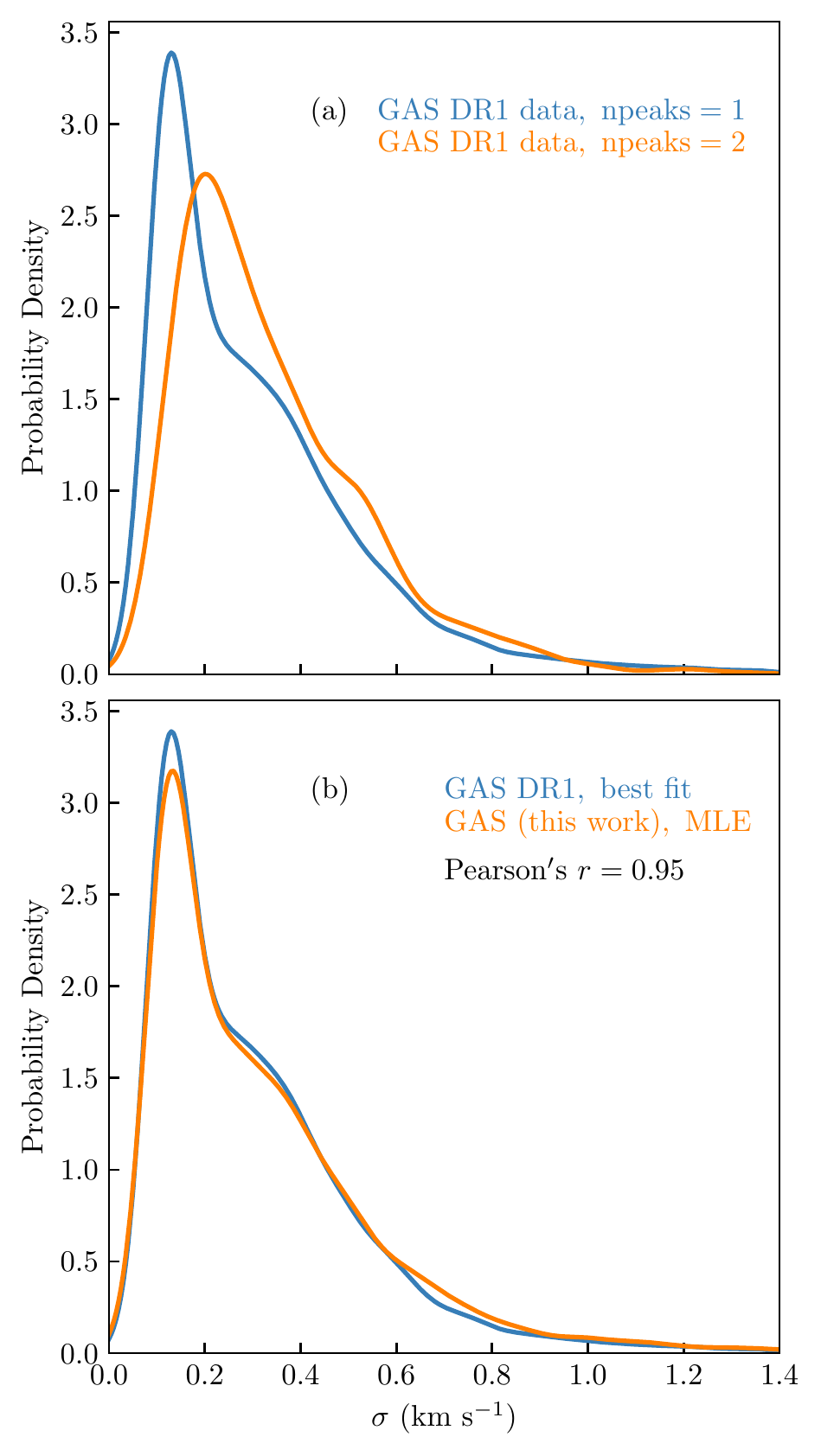}
    \caption{(a) The distribution of the GAS-DR1 best-fit velocity dispersion values, colored in blue and orange where this study finds one and two line-of-sight components, respectively. (b) Best-fit velocity dispersion distributions in GAS-DR1 results in blue and the corresponding MLE values from this work shown in orange.}
    \label{fig:hist}
\end{figure}

The original GAS-DR1 paper \citep{friesen+pineda+2017} shows the velocity dispersion distribution for all pixels in the regions studied. The distribution is composed of a substantial number of pixels with a narrow (subsonic) velocity dispersion and a wide distribution of broad velocity dispersion (corresponding to supersonic value). 
However, the degree to which the spectral multiplicity affected this broad velocity dispersion population was unclear. 
In the top panel of Figure~\ref{fig:hist} we plot the same KDE\footnote{SciPy implementation of Gaussian kernels with a bandwidth determined using Scott’s Rule} estimated velocity dispersion distributions with and without including the pixels where multiple components were found, and it shows that the DR1 reported values were biased towards larger values in places where we resolve two components. 
Meanwhile, the bottom panel of Figure~\ref{fig:hist} compares the velocity dispersion distributions from a common region where only one component is present. This shows good correspondence between the DR1 results and this study (Pearson's $r = 0.95$).
\added{Spectral multiplicity of \nnhp (1-0) has previously been reported in NGC 1333 \citep[cf. Fig. 12 in][]{hacar+2017}. While for all the regions reported to have a secondary component in \citeauthor{hacar+2017} we report at least one significant spectra with a secondary component, our method recovers more regions with two components.
We attribute the discrepancy to a mixture of difference in critical densities of the two transitions, unequal signal-to-noise coverage of the two data sets, and method-specific differences. Future studies that would perform nested sampling analysis on \citeauthor{hacar+2017} data should be able to pinpoint the reason for this apparent inconsistency.}

\subsection{Heuristical decisions on spectral multiplicity} \label{ssec:kcut}

The probabilistic approach in Bayesian model comparison does not implicitly require to make a decision on the number of line-of-sight components, but, 
while \Kij values alone can be used as a measure of model comparison, it is sometimes necessary to make an explicit decision for the preferred model. We adopt a decision threshold of $\ln \Kij = 5$, with higher \Kij-values indicating $\mathcal{M}_i$ model considered to be true. The chosen threshold
roughly corresponds to a ``decisive'' evidence strength on the Jeffreys' scale \citep{jeffreys1939}. To further justify our decision threshold, we have simulated a 64x64 synthetic spectral cube containing only the white noise component, and ran the same inference routine on it as on the actual GAS data. In total, 4096 spectra were generated with the noise amplitude randomly sampled from the DR1 RMS maps. We have found no false positives to be generated, with the maximum $\ln \Knilone$ value of 3.16 and 99.9th percentile of 2.13.
\added{Furthermore, a nested sampling analysis of M. Chen et al. (in prep.) on a control sample of 10,000 two-component synthetic spectra, generated to resemble the GAS data, finds no cases where two components are identified where only one was generated, and 12\% cases where one two components were mis-identified as one. We note that the 12\% fraction is dependent on the S/N ratio of the components and is dropping to zero when their separation becomes sufficiently large (M. Chen et al., in prep.).}
Following the approach above, we derive a map of spectral multiplicity for GAS results on NGC 1333.

Illustrating both the parameter estimation and model selection, Figure \ref{fig:specmaps} shows the spectral multiplicity with selected spectra overlaid with their best fit (MLE) profiles. The Bayes factors corresponding to detection of a single spectrum (\Knilone) and detection of a double component (\Ktwoone) are annotated alongside each spectrum. While the full Bayes factor and MLE maps are available online, we briefly describe a few representative spectra below. Ammonia spectra labelled as (1) are taken from a position where only one faint line components can be constrained, but no detection was reported in \cite{friesen+pineda+2017}. Note that the odds for finding a secondary component (\Ktwoone) are against the two-component model. The spectral lines denoted as (2) lie at the detection limit of the GAS DR1 results, but are significantly detected as one component ($\ln \Knilone = 10$) in our results. The map pointers (3) and (5) demonstrate a confident detection of secondary line components. Finally, a secondary component that would have normally missed the authors' by-eye inspection yet nonetheless is unambiguously present with a high certainty ($\ln \Ktwoone = 43$) is labelled as (3).

\subsection{Limitations and future prospects}

\begin{figure*}
    \centering
    \includegraphics[angle=90, width=0.55\textwidth]{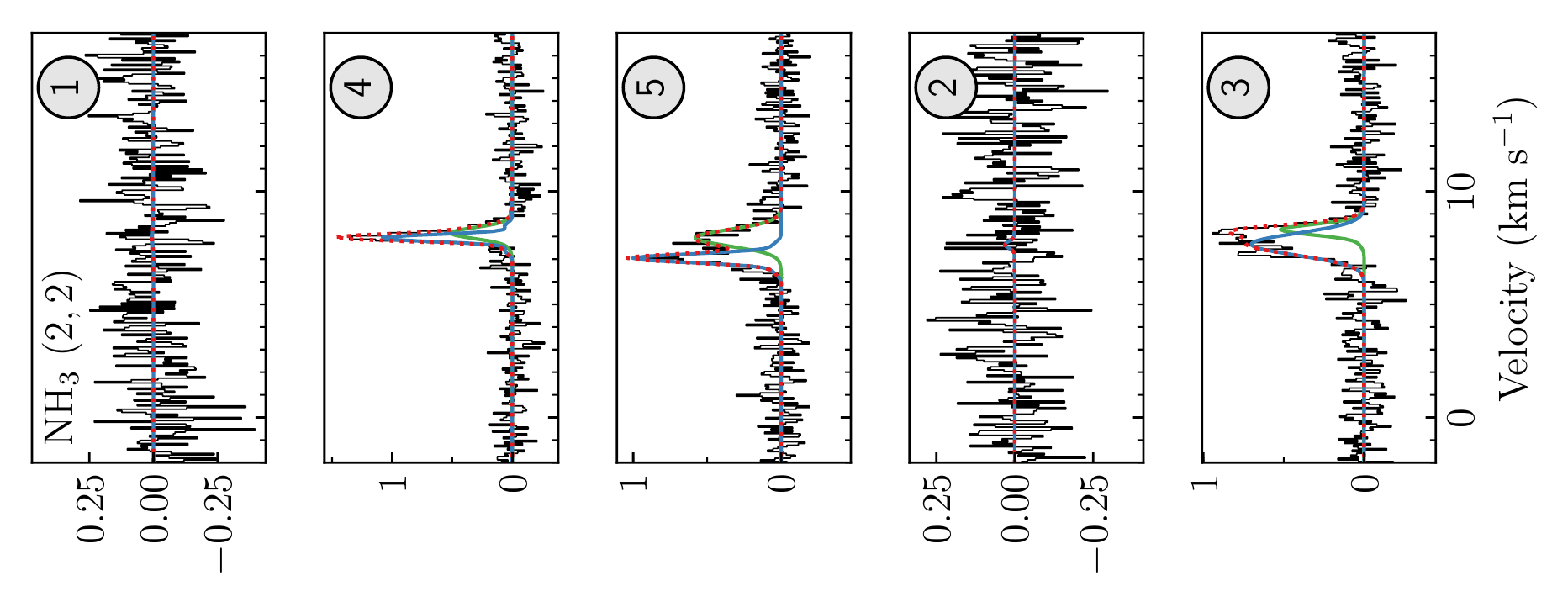}
    \includegraphics[angle=90, width=0.55\textwidth]{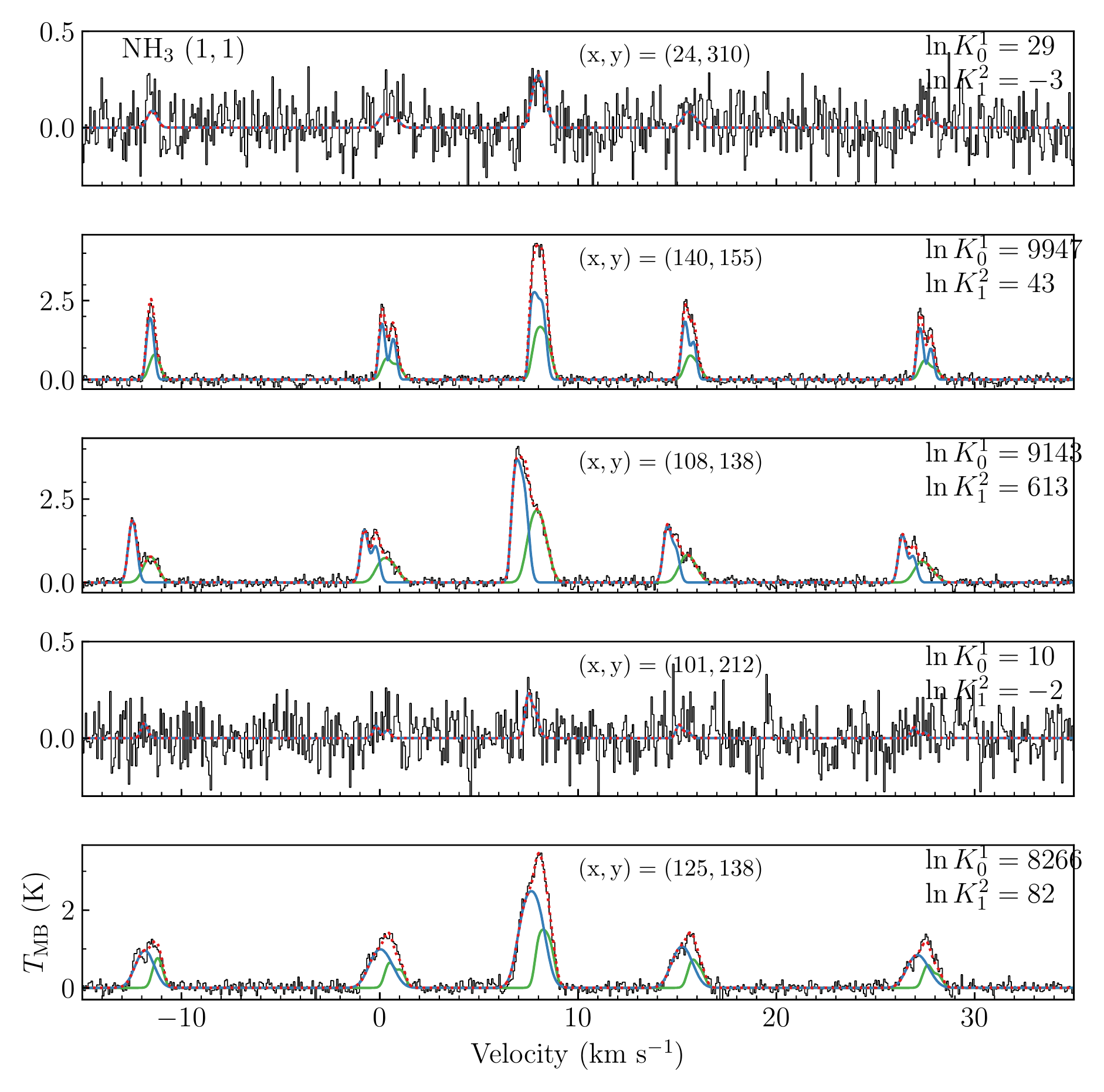}
    \includegraphics[angle=90, width=0.55\textwidth]{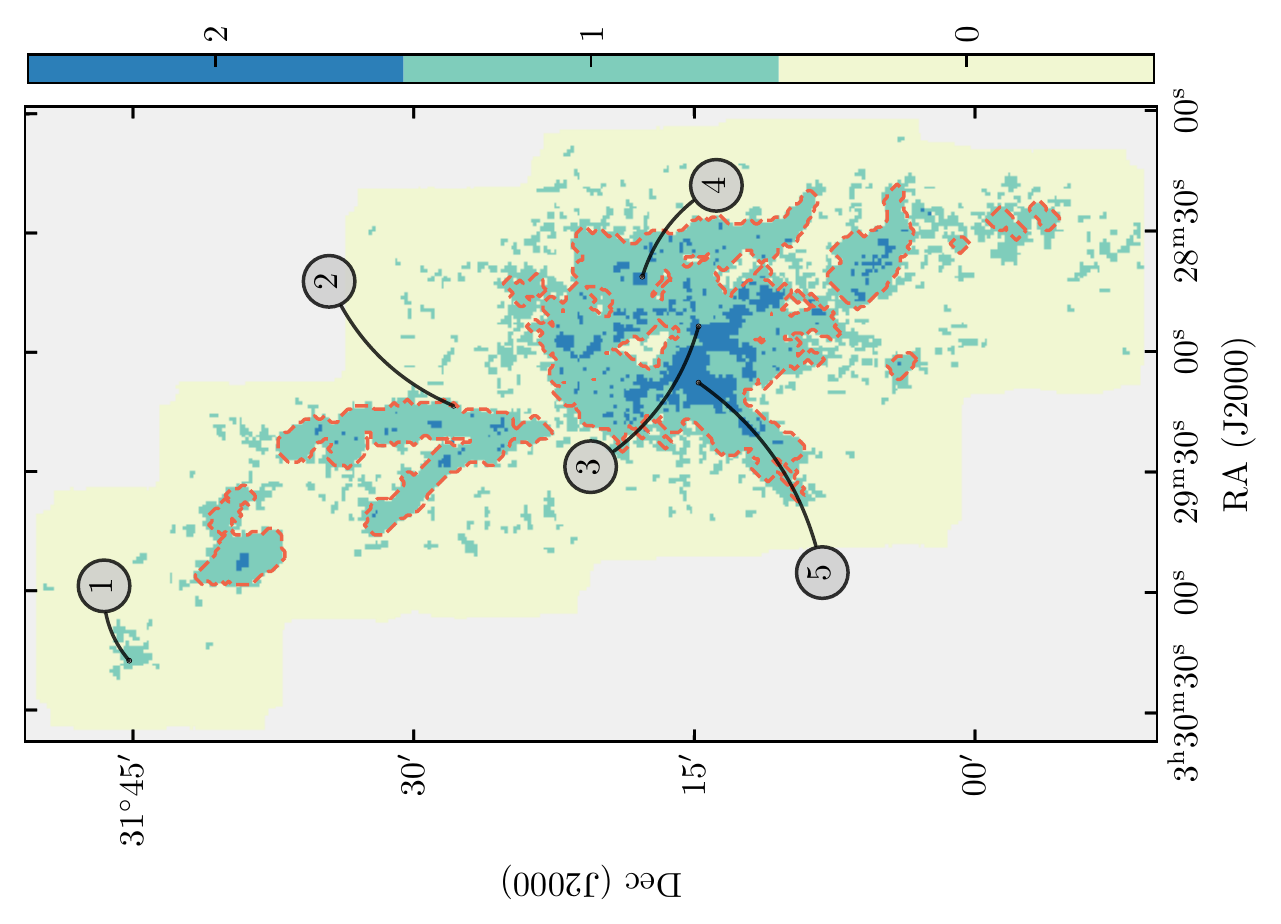}
    \caption{Selected spectra and the spectral multiplicity map. The map of spectral multiplicity ({\em left}) was constructed with a $\ln \Kij = 5$ threshold for selecting the more complex model. The numbered annotations point to selected spectra displayed on the {\em right} side. The \nhhh (1,1) and (2,2) spectra shown in the middle and right columns, respectively. The spectra are overlaid with their best MLE fits - solid colors for individual components, dotted red line for their total.}
    \label{fig:specmaps}
\end{figure*}

Despite the relative ease of the setup compared to conventional methods of modelling additional spectral components, the framework presented here must not be viewed as without limitations.
First and foremost, we emphasize that the choice of priors must be educated, and future empirical studies should be undertaken to fully validate the decision boundaries and the impact of prior volume on them. Nevertheless, we see the Bayesian model selection as an improvement over the state of the art heuristical quality control and by-eye human decisions for determining spectral multiplicity.
Secondly, the model selection framework we follow is only meaningful if the true model is included amongst the competing ones. As the only alternative to astrophysical ammonia emission included in the model space is Gaussian noise, the non-LTE effects, imaging artifacts, skewed spectral baselines, and other features not described by the emission models will often manifest as additional velocity components.
To illustrate this, we have sampled a three component model on all the five spectra shown on Fig. \ref{fig:specmaps}. We find that a third component is ruled out ($\ln \Kthreetwo < 2$) in all the spectra, except for (3), where three-component model is deemed the best ($\ln \Kthreetwo = 10.2; \ln \Kfourthree = -9.0$). We note that at 5 \kms, this extra feature is unlikely to be physical as its velocity is atypical for the cloud kinematics in the central NGC 1333 region (cf. Fig. \ref{fig:mlexoff}).
Thirdly, in the analysis above we operate under the assumption that the nearby spectra are not spatially-aware. We expect that an improved framework, where both the correlation between nearby spectra and the continuity of the physical properties are taken into the account, would greatly improve the performance of the Bayesian inference on spectral cube data.
Finally, the framework presented here is more limited by the computational time available than other approaches we compared it to. While sampling a moderately large spectral cube for up to two ammonia components is certainly feasible on a small cluster, sampling highly-dimensional models (e.g., fifteen or more free parameters) for large ALMA spectral cubes would require considerably more computational power.

\acknowledgments
{VS is grateful to Sebastian Grandis, Suhail Dhawan, and Linda Baronchelli for useful discussions on Bayesian statistics. The authors would like to thank Brian Svoboda for his insightful comments during the preparation of this manuscript.}

\facilities{GBT}

\software{MultiNest \citep{feroz+hobson2008, feroz+2009}, PyMultiNest \citep{buchner+2014}, scipy \citep{scipy}, astropy \citep{astropy2013, astropy2018}, pyspeckit \citep{ginsburg+mirocha2011}, APLpy \citep{aplpy}}

\bibliography{literature}

\end{document}